\documentclass[letterpaper,11pt]{article}

\usepackage[utf8]{inputenc}

\pdfoutput=1

\usepackage{amsmath,amssymb,amsthm,mathrsfs,bbm}
\usepackage{latexsym,amscd,amsbsy,amsfonts,dsfont}
\usepackage{hyperref}
\usepackage{graphicx,subcaption}

\numberwithin{equation}{section}

\begin{document}

\begin{titlepage}

\vspace{2cm}

\begin{center}
\renewcommand{\thefootnote}{\fnsymbol{footnote}}
{\Huge \bf Black objects in a five-dimensional swirling spacetime}
\vskip 32mm
{\large
{Adriano Vigan\`o\footnote{adriano.vigano@mi.infn.it}}
} \\

\renewcommand{\thefootnote}{\arabic{footnote}}
\setcounter{footnote}{0}
\vskip 10mm
\vspace{0.2 cm}
{\small \textit{Istituto Nazionale di Fisica Nucleare (INFN), Sezione di Milano \\
Via Celoria 16, I-20133 Milano, Italy}
}
\end{center}
\vspace{5.4 cm}

\begin{center}
{\bf Abstract}
\end{center}
We use the Ehlers symmetry of the five-dimensional Ernst equation to embed the Myers--Perry black hole and the rotating black ring in a five-dimensional swirling spacetime.
We show that both solutions can be made free of conical singularities, and that the swirl modifies the balance condition of the black ring.
We compute horizon areas, surface gravities and Komar charges, which satisfy a Smarr relation, and study the ergoregions.

\end{titlepage}

\tableofcontents

\newpage

\section{Introduction}

The study of exact solutions in higher-dimensional general relativity continues to gain considerable interest.
String theory, one of the most promising theories of quantum gravity, lives in more than four dimensions and within this theory the microscopic counting of the Bekenstein--Hawking entropy was achieved~\cite{Strominger:1996sh};
the AdS/CFT correspondence~\cite{Maldacena:1997re} relates the dynamics of a $d$-dimensional black hole with quantum field theory in $d-1$ dimensions;
the theoretical possibility of creating black holes in accelerators due to large extra dimensions has been proposed~\cite{Argyres:1998qn} in the past years.
These examples show the reasons why gravity in more than four dimensions is largely studied.
Moreover, gravity in higher dimensions is interesting in its own, because it sheds more light on the structure of the theory, and in particular on black holes, which are the most important Lorentzian manifolds in any dimensions.

The usual four-dimensional uniqueness theorems~\cite{Hawking:1971vc,Hawking:1973uf} do not hold in higher dimensions, and indeed there are plenty of black hole solutions with non-trivial topology:
Myers--Perry black holes~\cite{Myers:1986un}, with topology $S^3$;
black objects with extended horizons like black strings or $p$-branes;
black rings~\cite{Emparan:2001wn,Elvang:2004rt,Emparan:2006mm}, whose topology is $S^1\times S^2$;
the black Saturn~\cite{Elvang:2007rd}, that is a superposition of $S^3$ and $S^1\times S^2$;
black di-rings~\cite{Iguchi:2007is}, that are a superposition of two $S^1\times S^2$;
black holes with lens-space topology~\cite{Chen:2008fa}, whose topology is $S^3/\mathbb{Z}_2$;
and many others.
These examples show that there is a true zoology of black objects in higher-dimensional gravity:
in this sense, general relativity in $d>4$ is richer, but also more difficult because of the larger number of degrees of freedom.

Recently, a new class of four-dimensional spacetimes, dubbed \emph{swirling} universes, was discovered in~\cite{Astorino:2022aam}.
These solutions represent a universe (where a black hole can be embedded) characterized by a rotation that drags the observers contained within:
a sort of whirlpool generated by spacetime curvature.
The first example of black holes embedded in the swirling universe was given in~\cite{Astorino:2022aam}, in particular for the Schwarzschild and Kerr holes.
Subsequently, other swirling solutions have been built~\cite{Astorino:2022prj,Astorino:2023elf,Cisterna:2023uqf,Illy:2023iau,Barrientos:2024pkt,Barrientos:2024umq,Astorino:2025zse,DiPinto:2025yaa}, and many properties of these spacetimes have been investigated~\cite{Capobianco:2023kse,Capobianco:2024jhe,Capobianco:2025lxn,Moreira:2024sjq}.

Black holes in rotating five-dimensional backgrounds were previously found in minimal supergravity, for instance in the G\"odel universe~\cite{Gimon:2003ms,Herdeiro:2003un};
here, instead, the background is a vacuum solution without closed timelike curves.
The $SL(3,\mathbb{R})$ hidden symmetry of five-dimensional vacuum gravity with three commuting Killing vectors~\cite{Maison:1979kx} and its Ehlers subgroup have been used to generate rotating solutions~\cite{Giusto:2007fx,Suzuki:2024coe}.
After the first version of this work appeared, Ref.~\cite{Barrientos:2026swirl} constructed a two-parameter family of five-dimensional swirling backgrounds using the full coset formulation;
the background relevant for the black objects of Secs.~\ref{sec:myers-perry} and~\ref{sec:ring} is its single-swirl member.

Our aim is to extend the swirling universe to higher-dimensional gravity, and more specifically to the five-dimensional case.
In Sec.~\ref{sec:ernst}, we introduce the five-dimensional Ernst equation and discuss the exact Lie point symmetry that it enjoys.
Then, in Sec.~\ref{sec:swirl} we apply the Ehlers map to five-dimensional Minkowski spacetime, obtaining two swirling backgrounds:
the direct product of the four-dimensional swirling universe with a line, and a genuinely five-dimensional spacetime, which is the background of the black objects studied in the rest of the paper.
In Sec.~\ref{sec:myers-perry} we embed the Myers--Perry black hole in the five-dimensional swirling universe:
we prove that such a solution is free of conical singularities and we study its geometrical properties.
In Sec.~\ref{sec:ring} we embed the rotating black ring in the swirling universe:
in this case the solution is quite involved and some of its properties have to be investigated numerically.
Again, we prove that the solution can be made free of conical singularities, we show that the swirl modifies the balance condition of the ring, and then discuss the behaviour of its mass and angular momentum.
Finally, we sum up our findings and discuss possible future developments of the subject.

\section{Five-dimensional Ernst equation}
\label{sec:ernst}

We consider the five-dimensional Einstein equations in vacuum, $R_{\mu\nu}=0$, with a metric ansatz of the form
\begin{equation}
\label{eq:lwp-magnetic}
{ds}^2 = -\frac{\rho^2}{fh} {dt}^2 + \frac{f}{h} \bigl(d\psi-\omega dt\bigr)^2 + \frac{e^{2\gamma}}{fh} \bigl( {d\rho}^2+{dz}^2 \bigr) + h^2 {d\phi}^2 \,,
\end{equation}
where $f$, $h$, $\omega$ and $\rho$ are all functions of the coordinates $(\rho,z)$.
This ansatz represents a stationary and axisymmetric spacetime in Weyl coordinates, with possible rotation along the $\partial/\partial\psi$ direction, and it is characterized by three commuting Killing vectors:
$\partial/\partial t$, $\partial/\partial\psi$ and $\partial/\partial\phi$.

In~\cite{Yazadjiev:2008pt} it was shown that, for a metric of the form
\begin{equation}
\label{eq:lwp-electric}
{ds}^2 = - \frac{f}{h} \bigl(dt-\omega d\psi\bigr)^2 + \frac{\rho^2}{fh} {d\psi}^2 + \frac{e^{2\gamma}}{fh} \bigl( {d\rho}^2+{dz}^2 \bigr) + h^2 {d\phi}^2 \,,
\end{equation}
it is possible to rewrite the dimensionally reduced five-dimensional Einstein equations in the form of the Ernst equations
\begin{equation}
\label{eq:ernst-equation}
(\text{Re}\mathcal{E})\, \nabla^2\mathcal{E} = \vec{\nabla}\mathcal{E} \cdot \vec{\nabla}\mathcal{E} \,,
\end{equation}
where we introduced the 5D gravitational Ernst potential
\begin{equation}
\label{eq:ernst-potential}
\mathcal{E} = f + i\chi \,,
\end{equation}
and where the differential operators are expressed in standard three-dimensional cylindrical coordinates $(\rho,\varphi,z)$, with $\varphi$ an auxiliary azimuthal angle.
$\chi$ is the twist potential defined by
\begin{equation}
\label{eq:tiwst-potential}
\hat{\varphi} \times \vec{\nabla}\chi = \rho^{-1} f^2\, \vec{\nabla}\omega \,.
\end{equation}
The function $\gamma$ appearing in~\eqref{eq:lwp-electric}, which is left out by the definition of the Ernst potential, is found by quadratures via the equations
\begin{subequations}
\begin{align}
\partial_\rho\gamma & = \frac{\rho}{4(\text{Re}\mathcal{E})^2} \bigl( \partial_\rho\mathcal{E}\partial_\rho\mathcal{E}^* - \partial_z\mathcal{E}\partial_z\mathcal{E}^* \bigr)
+ \frac{3\rho}{4} \bigl[ (\partial_\rho\ln h)^2 - (\partial_z\ln h)^2 \bigr] \,, \\
\partial_z\gamma & = \frac{\rho}{2(\text{Re}\mathcal{E})^2}\, \text{Re}\bigl( \partial_\rho\mathcal{E}\partial_z\mathcal{E}^* \bigr)
+ \frac{3\rho}{2}\, \partial_\rho\ln h\, \partial_z\ln h \,,
\end{align}
\end{subequations}
where the last terms are the contribution of the function $h$.
Since $h$ is not affected by the transformation introduced below, $\gamma$ is invariant under it and can be read off directly from the seed.
The remarkable feature of the Ernst equation~\eqref{eq:ernst-equation} is that it reveals non-trivial symmetries of the equations of motion that are hidden in the usual five-dimensional formulation, as it happens in the four-dimensional case~\cite{Ernst:1967wx}.
In this sense, Eq.~\eqref{eq:ernst-equation} is not just a rewriting of the Einstein equations, but it is vital to generate transformations which give rise to solutions that are physically inequivalent to the seed ones (as to say, these maps are not simply gauge transformations).

Since Eq.~\eqref{eq:ernst-equation} is formally identical to the four-dimensional Ernst equation~\cite{Ernst:1967wx}, it enjoys the very same symmetries:
among the symmetries of the Ernst equation, the most interesting ones are the Lie point symmetries, controlled by continuous parameters.
In particular, Eq.~\eqref{eq:ernst-equation} is left invariant by the Ehlers map~\cite{Ehlers}
\begin{equation}
\label{eq:ehlers}
\mathcal{E}' = \frac{\mathcal{E}}{1+ij\mathcal{E}} \,,
\end{equation}
where $j\in\mathbb{R}$ is the arbitrary parameter of the transformation.
Since~\eqref{eq:ehlers} is an exact symmetry of the system, it transforms solutions of the vacuum Einstein equations into solutions of the vacuum Einstein equations.
In four dimensions, the Ehlers map~\eqref{eq:ehlers} applied to a stationary and axisymmetric metric (written in its two non-equivalent forms) generates a NUT parameter~\cite{ReinaTreves} or a swirling parameter~\cite{Astorino:2022aam}.

We can take advantage of this result and connect it to the fact that our ansatz~\eqref{eq:lwp-magnetic} can be obtained by double-Wick rotating the metric~\eqref{eq:lwp-electric} in the coordinates $t$ and $\psi$.
As is well known~\cite{Vigano:2022hrg}, the double-Wick rotation does not change the definition of the gravitational Ernst potential~\eqref{eq:ernst-potential} and of the twist potential~\eqref{eq:tiwst-potential}, and thus of the Ernst equation~\eqref{eq:ernst-equation}.
This means that the Ehlers map~\eqref{eq:ehlers} is still an exact symmetry of the Ernst system, and we can use it to map solutions into solutions.
By analogy with the four-dimensional case~\cite{Astorino:2022aam}, we expect that such a symmetry generates a swirling universe.

We stress that the formulation~\eqref{eq:lwp-magnetic}--\eqref{eq:tiwst-potential} requires $\partial/\partial\phi$ to be hypersurface-orthogonal, so that only the twist associated with $\partial/\partial\psi$ enters~\cite{Yazadjiev:2008pt}.
Equivalently, Eq.~\eqref{eq:ernst-equation} is the vacuum Ernst equation of the four-dimensional Einstein-frame metric obtained by Kaluza--Klein reduction along $\phi$, in which the dilaton $\ln h$ decouples and is left invariant by~\eqref{eq:ehlers}.
Hence the map~\eqref{eq:ehlers} switches on a single swirling parameter;
the full $SL(3,\mathbb{R})$ symmetry of five-dimensional vacuum gravity~\cite{Maison:1979kx,Giusto:2007fx} allows for a second, independent one, as recently shown in~\cite{Barrientos:2026swirl}.

What we have to do now is to choose a seed metric, cast it into the form~\eqref{eq:lwp-magnetic}, construct the Ernst potential~\eqref{eq:ernst-potential} and finally use the Ehlers map~\eqref{eq:ehlers} to generate a new solution of the Ernst equation~\eqref{eq:ernst-equation}.
Finally, we will reconstruct the metric using the definitions given above.
In the following, we will perform this computation by using three different seeds:
Minkowski spacetime (Sec.~\ref{sec:swirl}), Myers--Perry black hole (Sec.~\ref{sec:myers-perry}) and rotating black ring (Sec.~\ref{sec:ring}).

\section{Swirling universe}
\label{sec:swirl}

As a warm up, we apply the Ehlers map to five-dimensional Minkowski spacetime.
We first write it in cylindrical coordinates for a single rotation plane, with $\phi$ a Cartesian direction:
\begin{equation}
{ds}^2 = -{dt}^2 + {d\rho}^2 + {dz}^2 + \rho^2{d\psi}^2 + {d\phi}^2 \,,
\end{equation}
with the associated Ernst potential
\begin{equation}
\mathcal{E} = \rho^2 \,.
\end{equation}
From this, we construct the new Ernst potential
\begin{equation}
\mathcal{E}' = \frac{\rho^2}{1+ij\rho^2} \,,
\end{equation}
from which we read the twist and the angular velocity
\begin{equation}
\chi' = -j \frac{\rho^4}{1+j^2\rho^4} \implies
\omega' = 4 j z \,,
\end{equation}
and finally obtain the new metric, that is
\begin{equation}
\label{eq:swirling}
{ds}^2 = (1+j^2\rho^4) \bigl( -{dt}^2 + {d\rho}^2 + {dz}^2 \bigr)
+ \frac{\rho^2}{1+j^2\rho^4} (d\psi - 4jz\, dt)^2 + {d\phi}^2 \,,
\end{equation}
where $j$ is the swirling parameter.
The metric is the direct product of the four-dimensional swirling universe~\cite{Astorino:2022aam} with the flat direction $\phi$.
There are three Killing vectors, $\partial/\partial t$, $\partial/\partial\psi$ and $\partial/\partial\phi$:
the first two are related to stationarity and to axisymmetry with respect to the $\psi$-plane, while $\partial/\partial\phi$ generates translations along the flat direction.

The spacetime is naturally free of conical singularities, and the Kretschmann scalar is
\begin{equation}
\label{eq:kretch}
R_{\mu\nu\rho\sigma} R^{\mu\nu\rho\sigma} = 192 j^2 \frac{j^6\rho^{12}-15j^4\rho^8+15j^2\rho^4-1}{(1+j^2\rho^4)^6} \,.
\end{equation}
We see that the scalar invariant~\eqref{eq:kretch} is everywhere well-behaved, so there are no curvature singularities.
As $\rho\to\infty$, we notice that $R_{\mu\nu\rho\sigma} R^{\mu\nu\rho\sigma}\sim\frac{192}{j^4\rho^{12}}$.

The angular velocity is given by
\begin{equation}
\Omega = -\frac{g_{t\psi}}{g_{\psi\psi}} = 4jz \,,
\end{equation}
the very same expression as in the four-dimensional case~\cite{Astorino:2022aam}.
Thus, the angular velocity of the spacetime changes sign by crossing the origin of the $z$-axis and increases as the absolute value of $z$ grows.
We expect a frame dragging for particles in such a spacetime, as can be verified by inspecting the geodesics:
we briefly analyze the geodesics of the spacetime~\eqref{eq:swirling} by relying on the methods of~\cite{Capobianco:2023kse}.

The geodesics are defined as solutions to the equations
\begin{equation}
\ddot{x}^\mu + \Gamma^\mu_{\;\nu\sigma} \dot{x}^\nu \dot{x}^\sigma = 0 \,,
\end{equation}
we introduced $\dot{x}^\mu\equiv\frac{dx^\mu}{d\tau}$, $\tau$ is an affine parameter and $x^\mu=\bigl(t(\tau),\rho(\tau),z(\tau),\psi(\tau),\phi(\tau)\bigr)$.
A first order formulation of the geodesic equations is given by the Hamilton--Jacobi equation
\begin{equation}
2 \frac{\partial S}{\partial\tau} = g^{\mu\nu} \partial_\mu S\, \partial_\nu S \,,
\end{equation}
where $S$ is the Hamilton principal function.

We take advantage of the symmetries of the swirling universe, and define the local conserved quantities
\begin{equation}
-E = g_{tt}\dot{t} + g_{t\psi}\dot{\psi} \,, \quad
J_\psi = g_{\psi\psi}\dot{\psi} + g_{t\psi}\dot{t} \,, \quad
J_\phi = g_{\phi\phi} \dot{\phi} \,,
\end{equation}
plus a fourth conserved quantity defined by the rest mass
\begin{equation}
\epsilon = g_{\mu\nu} \dot{x}^\mu\dot{x}^\nu \,,
\end{equation}
where $\epsilon=-1$ for timelike geodesics and $\epsilon=0$ for null geodesics, respectively.
We choose a separability ansatz for the Hamilton principal function
\begin{equation}
S = \frac{1}{2}\epsilon\tau - Et + J_\psi \psi + J_\phi \phi + S_\rho(\rho) + S_z(z) \,.
\end{equation}
The resulting Hamilton--Jacobi equation is separable, and gives rise to the equations
\begin{subequations}
\begin{align}
(\partial_\rho S_\rho)^2 & = \frac{\rho^2[k+(\epsilon-J_\phi^2)F]-J_\psi^2F^2}{\rho^2} \,, \\
(\partial_z S_z)^2 & = E^2 - k + J_\psi\omega (J_\psi\omega-2E) \,,
\end{align}
\end{subequations}
where $F\equiv1+j^2\rho^4$, $\omega=4jz$, and $k$ is the separation constant.

\begin{figure}
\centering
\begin{subfigure}{0.45\textwidth}
    \centering
    \includegraphics[scale=0.7]{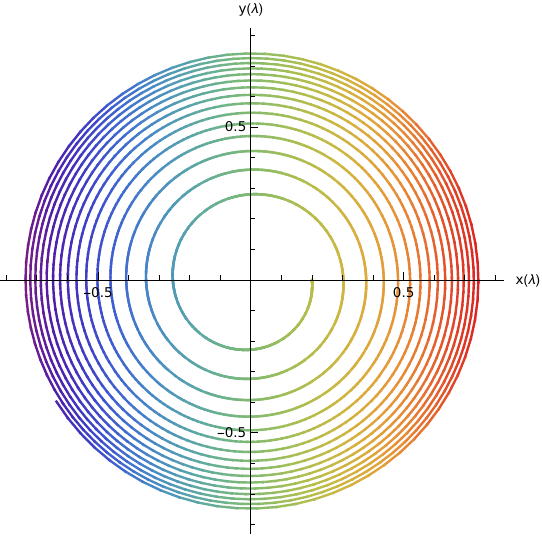}
    \caption{Projection of the geodesic motion on the $xy$ plane.}
    \label{fig:swirl-geo1}
\end{subfigure}
\hfill
\medskip
\begin{subfigure}{0.45\textwidth}
    \centering
    \includegraphics[scale=0.7]{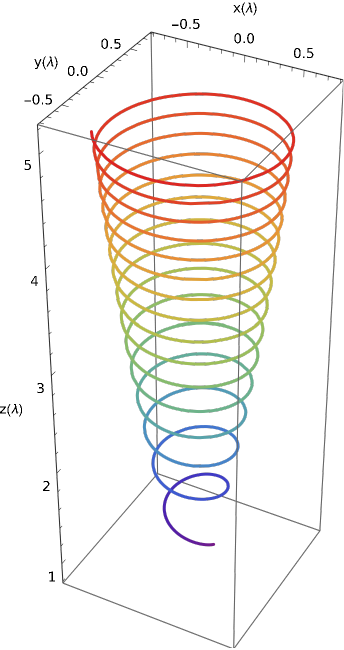}
    \caption{Geodesic motion in the $xyz$ space.}
    \label{fig:swirl-geo2}
\end{subfigure}
\caption{Plots showing the motion of a massless particle in the swirling universe~\eqref{eq:swirling}, with parameters $j=2$, $E=5$, $J_\psi=0.4$, $J_\phi=0$, $\epsilon=0$, $k=5$.
We observe a frame dragging due to the presence of the angular velocity parametrized by $j$.}
\label{fig:swirl-geoMassless}
\end{figure}

\begin{figure}
\centering
\begin{subfigure}{0.45\textwidth}
    \centering
    \includegraphics[scale=0.7]{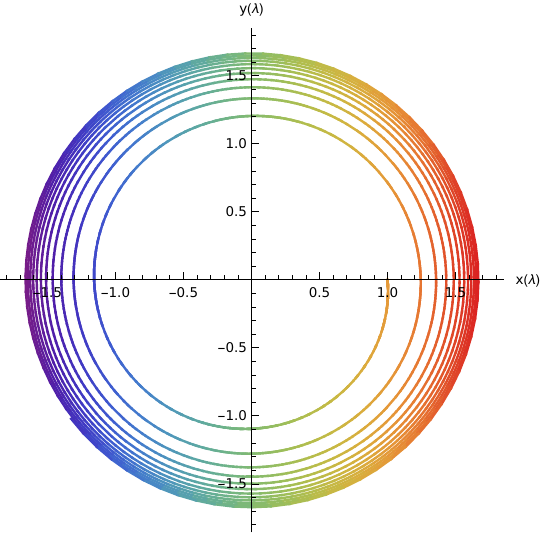}
    \caption{Projection of the geodesic motion on the $xy$ plane.}
    \label{fig:swirl-geo3}
\end{subfigure}
\hfill
\medskip
\begin{subfigure}{0.45\textwidth}
    \centering
    \includegraphics[scale=0.7]{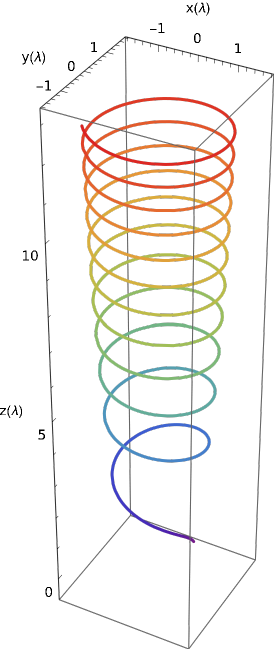}
    \caption{Geodesic motion in the $xyz$ space.}
    \label{fig:swirl-geo4}
\end{subfigure}
\caption{Plots showing the motion of a massive particle in the swirling universe~\eqref{eq:swirling}, with parameters $j=0.22$, $E=5$, $J_\psi=1$, $J_\phi=0$, $\epsilon=-1$, $k=2.2$.
We observe a frame dragging due to the presence of the angular velocity parametrized by $j$.}
\label{fig:swirl-geoMassive}
\end{figure}

The geodesic equations are finally given by the generalized momenta $g_{\mu\nu}\dot{x}^\nu=\partial_\mu S$, and they read
\begin{subequations}
\label{eq:geodesics}
\begin{align}
\frac{dt}{d\lambda} & = E - J_\psi \omega \,, \\
\frac{d\rho}{d\lambda} & = \pm \frac{1}{\rho} \sqrt{\rho^2[k+(\epsilon-J_\phi^2)F]-J_\psi^2F^2} \,, \\
\frac{dz}{d\lambda} & = \pm \sqrt{E^2 - k + J_\psi\omega (J_\psi\omega-2E)} \,, \\
\frac{d\psi}{d\lambda} & = \frac{F^2}{\rho^2} J_\psi + \omega(E-\omega J_\psi) \,, \\
\frac{d\phi}{d\lambda} & = F J_\phi \,,
\end{align}
\end{subequations}
where we introduced $d\tau=Fd\lambda$.

Eqs.~\eqref{eq:geodesics} are formally identical to those of~\cite{Capobianco:2023kse}, and thus one can integrate them to find a complete set of solutions.
We are interested in showing the qualitative behavior of particles in the universe~\eqref{eq:swirling}, thus we numerically integrate the equations and plot the results\footnote{To plot the geodesics, we introduced the Cartesian coordinates $x=\rho\cos\psi$ and $y=\rho\sin\psi$.} in Fig.~\ref{fig:swirl-geoMassless} and~\ref{fig:swirl-geoMassive}.
In both cases, we observe that the universe rotation, due to the presence of $j$, drags the massless and massive particles, which whirl around the $z$-axis.
This is expected:
since $h=1$, the spacetime~\eqref{eq:swirling} is the direct product of the four-dimensional swirling universe with a line, so that its curvature~\eqref{eq:kretch} and its geodesics with $J_\phi=0$ coincide with the four-dimensional ones~\cite{Astorino:2022aam,Capobianco:2023kse}.
It is therefore the natural background for objects with Kaluza--Klein asymptotics.

The black objects of Secs.~\ref{sec:myers-perry} and~\ref{sec:ring}, whose seeds are asymptotically $\mathbb{R}^{1,4}$, live instead in the background obtained by applying~\eqref{eq:ehlers} to Minkowski spacetime written in terms of two rotation planes, i.e.~the $m\to0$ limit of the solution~\eqref{eq:tang-swirling} below:
\begin{equation}
\label{eq:swirl-axial}
{ds}^2 = \Upsilon \bigl( -{dt}^2 + {dr_1}^2 + {dr_2}^2 \bigr)
+ \frac{r_1^2}{\Upsilon} \bigl[ d\psi - j(2r_2^2-r_1^2)\, dt \bigr]^2 + r_2^2 {d\phi}^2 \,, \qquad
\Upsilon = 1 + j^2 r_1^4 r_2^2 \,,
\end{equation}
where $r_1$ and $r_2$ are the radii in the $\psi$- and $\phi$-planes, respectively.
This is a genuinely five-dimensional \emph{swirling universe}.
It is regular at the origin and on both axes, with $\Delta\psi=\Delta\phi=2\pi$, and it has no closed timelike curves, since $g^{tt}=-1/\Upsilon<0$.
The frame dragging reverses its sign on the cone $r_1=\sqrt{2}\,r_2$.
Contrary to~\eqref{eq:swirling}, however, its curvature is not bounded:
on the two axes one finds
\begin{equation}
\label{eq:kretch-axial}
R_{\mu\nu\rho\sigma} R^{\mu\nu\rho\sigma}\big|_{r_1=0} = -192 j^2 r_2^2 \,, \qquad
R_{\mu\nu\rho\sigma} R^{\mu\nu\rho\sigma}\big|_{r_2=0} = 72 j^4 r_1^8 - 192 j^2 r_1^2 \,,
\end{equation}
which diverge at large distance along the axes, even though the Kretschmann scalar decays at fixed angle.
The spacetime~\eqref{eq:swirl-axial} is the single-swirl member of the two-parameter family of Ref.~\cite{Barrientos:2026swirl}, where its asymptotic structure is analysed in detail.
In the following sections, we will embed black objects in the background~\eqref{eq:swirl-axial}, and we will see that, despite its resemblance with the four-dimensional case, some peculiar features related to the higher dimension will appear.

\section{Swirling Myers--Perry}
\label{sec:myers-perry}

The first black object that we embed in the swirling spacetime~\eqref{eq:swirl-axial} is the Myers--Perry black hole~\cite{Myers:1986un}.
Such a solution was constructed for arbitrary dimensions $D$ and represents a rotating black hole with $\lfloor\frac{D-1}{2}\rfloor$ independent angular momenta, one for each rotational Killing field (see also~\cite{Emparan:2008eg} for a thorough review).

Here, we limit ourselves to the $D=5$ case with only one angular momentum:
the metric is then
\begin{equation}
\label{eq:mp}
{ds}^2 = -{dt}^2 + \frac{m}{\Sigma} \bigl( dt - a\sin^2\theta {d\psi} \bigr)^2 + \Sigma \biggl( \frac{{dr}^2}{\Delta} +{d\theta}^2 \biggr) + (r^2+a^2)\sin^2\theta {d\psi}^2 + r^2\cos^2\theta {d\phi}^2 \,,
\end{equation}
where we defined
\begin{equation}
\Sigma = r^2 + a^2\cos^2\theta \,, \qquad
\Delta = r^2 - m + a^2 \,.
\end{equation}
The angular coordinates are defined in the following ranges:
$\theta\in[0,\frac{\pi}{2}]$, $\psi\in[0,2\pi]$, $\phi\in[0,2\pi]$.
The solution has three Killing fields, $\partial/\partial t$, $\partial/\partial\psi$ and $\partial/\partial\phi$.
The parameters $m$ and $a$ are related to the mass and the angular momentum with respect to the rotational axis $\partial/\partial\psi$ of the black hole.
The horizon has topology $S^3$ and is defined by the condition $\Delta=0$, that gives $r_H=\sqrt{m-a^2}$:
it exists only if the condition $m-a^2>0$ is satisfied.
The black hole becomes extremal when $m=a^2$, but in this regime it has zero area and reduces to a naked ring singularity~\cite{Myers:1986un}.
The horizons are Killing horizons with respect to the Killing vector $\partial_t+\Omega_H\partial_\psi$, where $\Omega_H=a/m$ is the angular velocity along the vector $\partial/\partial\psi$, evaluated at the horizon $r_H$.

We compare the Myers--Perry metric~\eqref{eq:mp} to the ansatz~\eqref{eq:lwp-magnetic} in order to recognize the Ernst potential:
by defining
\begin{equation}
\rho = r\sin\theta\cos\theta \sqrt{\Delta} \,, \qquad
z = \frac{r^2-(r^2+\Delta)\cos^2\theta}{2} \,,
\end{equation}
we can construct the Ernst potential~\eqref{eq:ernst-potential} and apply the Ehlers transformation~\eqref{eq:ehlers} to the seed metric~\eqref{eq:mp}.
We obtain
\begin{equation}
\label{eq:mp-swirling}
{ds}^2 = \Lambda \bigl(d\psi-\omega dt\bigr)^2 + \Lambda^{-1} \biggl[ -\frac{\rho^2}{h^2} {dt}^2 + \Xi \sin^2\theta \biggl( \frac{{dr}^2}{\Delta} + {d\theta}^2 \biggr) \biggr] + h^2 {d\phi}^2 \,,
\end{equation}
where
\begin{equation}
\Lambda^{-1} = \chi_{(0)} + j \chi_{(1)} + j^2 \chi_{(2)} \,, \qquad
\omega = \omega_{(0)} + j \omega_{(1)} + j^2 \omega_{(2)} + \Omega_0 \,.
\end{equation}
We define the functions
\begin{equation}
h = r \cos\theta \,, \qquad
\rho = h \sin\theta \sqrt{\Delta} \,, \qquad
\Xi = \Sigma\Delta + m(r^2+a^2) \,,
\end{equation}
and
\begin{subequations}
\begin{align}
\chi_{(0)} & = \frac{\Sigma}{\Xi\sin^2\theta} \,, \\
\chi_{(1)} & = \frac{2am\cot^2\theta[a^2+r^2(2-\cos^2\theta)]}{\Xi} \,, \\
\begin{split}
\chi_{(2)} & = \frac{\cot^2\theta}{\Xi} \Bigl\{
r^2 (r^6+a^6\cos^2\theta) \sin^4\theta \\
&\quad + a^2r^2 \bigl[ m^2\cos^2\theta(2-\cos^2\theta)^2 + 2mr^2\sin^6\theta + r^4(2-3\cos^2\theta+\cos^6\theta) \bigr] \\
&\quad + a^4 \bigl[ m^2 + 2mr^2\sin^6\theta + r^4 (1-3\cos^4\theta+2\cos^6\theta) \bigr]
\Bigr\}
\,,
\end{split}
\\
\omega_{(0)} & = \frac{am}{\Xi} \,, \\
\begin{split}
\omega_{(1)} & = \frac{1}{a^2\Xi}
\Bigl\{
\bigl[ 2m(a^2+m)-a^2r^2 \bigr] \bigl[ r^4+a^2(r^2+m) \bigr] \\
&\quad + a^2\Delta \bigl[ m(a^2+2m)+3r^4 \bigr] \cos^2\theta
+ a^4\Delta (3r^2+a^2-m) \cos^4\theta \Bigr\} \,,
\end{split}
\\
\begin{split}
\omega_{(2)} & = \frac{m}{a^5\Xi}
\Bigl\{
m (a^6+a^4m-3a^2m^2+5m^3) \bigl[ r^4+a^2(r^2+m)+a^2\Delta \cos^2\theta \bigr] \\
&\quad + a^6\Delta \bigl[ a^4+3r^4+a^2(3r^2-m)-r^4\cos^2\theta \bigr] \cos^4\theta
\Bigr\} \,,
\end{split}
\end{align}
\end{subequations}
with constant
\begin{equation}
\label{eq:mp-omega0}
\Omega_0 =-\frac{jm^2}{a^5} \bigl(2a^3 + a^4jm - 3a^2jm^2 + 5jm^3\bigr) \,.
\end{equation}
The field equations determine $\omega$ only up to an additive constant, i.e.~up to the choice of the stationary Killing vector among the combinations $\partial_t+c\,\partial_\psi$.
The constant $\Omega_0$ removes the terms that are singular as $a\to0$ and fixes this freedom:
for this choice, the value of $\omega$ on the horizon is
\begin{equation}
\label{eq:omega-horizon}
\omega\big|_{r=r_H} = \frac{(a+jm^2)(1+ajm)}{m} \,.
\end{equation}
Since the spacetime is not asymptotically flat, there is no preferred choice:
the values of the mass and of the angular velocity computed in Sec.~\ref{sec:mp-smarr}, and the shape of the ergoregions of Sec.~\ref{sec:mp-ergo}, refer to this frame.

By setting $j=0$, we recover the Myers--Perry solution~\eqref{eq:mp}, while, when $a=0$ we find the swirling Tangherlini black hole:
\begin{equation}
\label{eq:tang-swirling}
{ds}^2 = \frac{r^2\sin^2\theta}{\Upsilon} \bigl(d\psi-\omega dt\bigr)^2 + \Upsilon \biggl[ -\biggl(1-\frac{m}{r^2}\biggr) {dt}^2 + \frac{{dr}^2}{1-\frac{m}{r^2}} + r^2 {d\theta}^2 \biggr] + r^2\cos^2\theta {d\phi}^2 \,,
\end{equation}
where in this subcase the functions reduce to
\begin{equation}
\label{eq:tang-functions}
\Upsilon = 1 + j^2 r^6 \cos^2\theta \sin^4\theta \,, \qquad
\omega = j \bigl[ 3(r^2-m)\cos^2\theta - r^2 + 2m \bigr] \,.
\end{equation}
For $m=0$, and in terms of $r_1=r\sin\theta$ and $r_2=r\cos\theta$, the metric~\eqref{eq:tang-swirling} reduces to the background~\eqref{eq:swirl-axial}.

Going back to the swirling Myers--Perry~\eqref{eq:mp-swirling}, the position of the horizon is left unchanged by the new background:
it is still given by $\Delta=0$, i.e.~the horizon is $r_H=\sqrt{m-a^2}$.
The topology of the event horizon is $S^3$, even if deformed, and it rotates with respect to the axis of the Killing field $\partial/\partial\psi$.
Outside the event horizon there are no curvature singularities at finite distance:
inspection of the Kretschmann scalar $R_{\mu\nu\rho\sigma}R^{\mu\nu\rho\sigma}$ reveals that the curvature singularity is located at $r=0$ and $\theta=\pi/2$, as usual.
However, as for the background~\eqref{eq:swirl-axial}, the curvature is unbounded along the axes at large distance.
For instance, for $a=0$ one finds on the two axes
\begin{equation}
R_{\mu\nu\rho\sigma}R^{\mu\nu\rho\sigma}\big|_{\theta=0} = \frac{72m^2}{r^8} - 192j^2r^2 \,, \qquad
R_{\mu\nu\rho\sigma}R^{\mu\nu\rho\sigma}\big|_{\theta=\pi/2} = \frac{72m^2}{r^8} + 336j^2m - 192j^2r^2 + 72j^4r^8 \,,
\end{equation}
which diverge as $r\to\infty$.
This is in contrast with the bounded curvature~\eqref{eq:kretch} of the four-dimensional swirling universe.

\subsection{Regularity}

The solution must be elementarily flat in the vicinity of the axes of the Killing fields $\partial/\partial\phi$  and $\partial/\partial\psi$ located at $\theta=\pi/2$ and $\theta=0$, respectively, to avoid conical singularities.
It is convenient to perform the computation by defining $x=\cos\theta$, so that we have to consider the limits $x\to0$ and $x\to1$.

The $x\phi$ sector for $x\to0$ is conformal to
\begin{equation}
{ds}_{x\phi}^2 \propto {dx}^2 + x^2 {d\phi}^2 \,,
\end{equation}
thus the axis is regular, which means that the periodicity of the angular coordinate $\phi$ is simply $\Delta\phi=2\pi$.

On the other hand, the $x\psi$ sector for $x\to1$ is conformal to
\begin{equation}
{ds}_{x\psi}^2 \propto \frac{(1+ajm)^2}{2s} {ds}^2 + \frac{2s}{(1+ajm)^2} {d\psi}^2 \,,
\end{equation}
where we defined $s=1-x$.
By introducing the new coordinate $u^2=\frac{2s}{(1+ajm)^2}$, we get
\begin{equation}
{ds}_{x\psi}^2 \propto {du}^2 + \frac{u^2}{(1+ajm)^4} {d\psi}^2 \,.
\end{equation}
Thus, to make the axis regular and avoid conical singularity, we require that the periodicity of the coordinate $\psi$ is
\begin{equation}
\label{eq:conic-mp}
\Delta\psi = 2\pi (1+ajm)^2 \,.
\end{equation}
With this choice, the solution is free of conical singularities.
This is a notable difference with respect to the four-dimensional case~\cite{Astorino:2022aam,DiPinto:2025yaa}, where the Kerr solution in a swirling universe is affected by the presence of conical defects.
The difference has a simple origin:
in four dimensions both semi-infinite axes belong to the same Killing vector and the twist potential jumps across the horizon, so that a single periodicity cannot regularize both of them;
in five dimensions, instead, the two semi-infinite axes belong to $\partial/\partial\psi$ and $\partial/\partial\phi$, whose periodicities are independent.
Assigning to the periodicity of $\psi$ the value~\eqref{eq:conic-mp}, on the other hand, we obtain a black hole solution that is free of conical singularities and of curvature singularities at finite distance outside the event horizon.
In the following, we restrict to $1+ajm>0$:
for $ajm=-1$, both the periodicity~\eqref{eq:conic-mp} and the horizon area~\eqref{eq:mp-area} vanish.

The periodicity~\eqref{eq:conic-mp} reduces to $2\pi$ whenever one of the three parameters is zero:
when $j=0$ we recover the Myers--Perry solution, when $a=0$ we recover the swirling Tangherlini spacetime and, finally, when $m=0$ we find the swirling background~\eqref{eq:swirl-axial} (the parameter $a$ can be reabsorbed by a redefinition of the coordinates).
All of these spacetimes are thus regular and do not need any redefinition of the periodicity.

\subsection{Geometry and Smarr law}
\label{sec:mp-smarr}

We are now interested in studying the geometrical properties of the solution~\eqref{eq:mp-swirling}, and their relation with black hole thermodynamics.

We define the surface gravity on the horizon in the usual way
\begin{equation}
\label{eq:kappa-def}
\kappa_H = \sqrt{-\frac{1}{2} \nabla_\mu\zeta_\nu \nabla^\mu\zeta^\nu} \,,
\end{equation}
where $\zeta^\mu$ is the generator of the Killing horizon.
Since the period of $\psi$ is~\eqref{eq:conic-mp}, it is convenient to introduce the angle $\tilde\psi=2\pi\psi/\Delta\psi$, whose period is $2\pi$.
The generator is then
\begin{equation}
\label{eq:mp-zeta}
\zeta = \frac{\partial}{\partial t} + \omega\big|_{r=r_H} \frac{\partial}{\partial\psi}
= \frac{\partial}{\partial t} + \Omega_H \frac{\partial}{\partial\tilde\psi} \,,
\end{equation}
where we introduced the angular velocity of the horizon
\begin{equation}
\Omega_H = \frac{2\pi}{\Delta\psi}\, \omega \biggr|_{r=r_H}
= \frac{a+jm^2}{m(1+ajm)} \,.
\end{equation}
Notice the presence of the periodicity of the azimuthal angle $\psi$.
Once explicitly evaluated, the surface gravity~\eqref{eq:kappa-def} gives
\begin{equation}
\label{eq:mp-kappa}
\kappa_H = \frac{r_H}{m} \,,
\end{equation}
which is the same result as for the asymptotically flat Myers--Perry black hole~\cite{Myers:1986un}.
Actually, the surface gravity is defined up to a constant scale factor, since in this case the normalization of the Killing vector at infinity is not clear \emph{a priori}.
We choose, however, the same normalization as in the Minkowski case to retain a well-defined limit for the case $j=0$.

The horizon area of the solution is found by restricting the metric on the horizon defined by $r=r_H$ and $t=$ const
\begin{equation}
\label{eq:mp-area}
\mathcal{A}_H = \int_H d\theta\, d\psi\, d\phi \sqrt{g_{\theta\theta}g_{\psi\psi}g_{\phi\phi}}
= 2\pi^2 m \, r_H (1+ajm)^2 \,.
\end{equation}
We observe that the presence of the swirling background modifies the area of the horizon by the factor $\Delta\psi/2\pi=(1+ajm)^2$, as for the angular velocity.

The surface gravity and the horizon area have a thermodynamical representation:
they are related to the temperature and the entropy of the black hole, respectively, by $T_H=\kappa_H/2\pi$ and $S=\mathcal{A}_H/4$ (we set $G=1$).

Since the swirling Myers--Perry black hole is not asymptotically flat, we cannot evaluate the ADM mass and angular momentum at infinity~\cite{Abdolrahimi:2014qja}.
However, we can consistently define \emph{local} mass and angular momentum by means of Komar integrals~\cite{Komar:1958wp} evaluated on the horizon.
Let $\xi=\partial/\partial t$ and $\eta=\partial/\partial\tilde\psi=(\Delta\psi/2\pi)\,\partial/\partial\psi$ be the timelike and spacelike Killing vectors, respectively, associated with time translations and rotations in the $\psi$ direction.
Then the local mass and the local angular momentum on the horizon are defined as
\begin{equation}
M_H = -\frac{3}{32\pi} \int_H dS^{\mu\nu} \nabla_\mu\xi_\nu \,, \qquad
J_H = \frac{1}{16\pi} \int_H dS^{\mu\nu} \nabla_\mu\eta_\nu \,.
\end{equation}
The surface element is $dS^{\mu\nu}= \sqrt{\sigma}\,r^{[\mu} n^{\nu]}$, where $r^\mu$ and $n^\mu$ are the unit vectors orthogonal to the $r=$ const and $t=$ const hypersurfaces, respectively, and $\sigma$ is the determinant of the $t,r=$ const part of the metric.
Computation of the Komar integrals gives
\begin{equation}
\label{eq:mp-komar}
M_H = \frac{3\pi}{8} m (1+ajm)^3 \,, \qquad
J_H = \frac{\pi}{4} am (1+ajm)^3 \,.
\end{equation}
We immediately observe that turning off the swirling parameter, $j=0$, gives the Myers--Perry charges $M=\frac{3\pi}{8}m$ and $J=\frac{\pi}{4}am$, as one expects.
We notice that the swirling and the rotation parameters couple to the mass parameter and modify the conserved charges by the factor $(1+ajm)^3$.
For $ajm>0$ both the mass and the angular momentum are larger with respect to the asymptotically flat case, while for $-1<ajm<0$ they are smaller.

The presence of the swirling parameter $j$ alone is not sufficient to generate a non-zero angular momentum:
if $a=0$, the total angular momentum~\eqref{eq:mp-komar} is zero irrespectively of the value of $j$.
This observation is consistent with the four-dimensional case~\cite{Astorino:2022aam}:
the Komar integral represents a monopole charge, thus the swirling parameter seems to generate a ``dipole'' rotation.

We see that the Komar charges~\eqref{eq:mp-komar}, and the geometrical quantities~\eqref{eq:mp-kappa} and~\eqref{eq:mp-area}, satisfy a local Smarr law~\cite{Smarr:1972kt}
\begin{equation}
\frac{2}{3} M_H = \frac{\kappa_H}{8\pi} \mathcal{A}_H + \Omega_H J_H \,,
\end{equation}
that coincides with the Smarr relation for the asymptotically flat Myers--Perry black hole~\cite{Myers:1986un}.
Since the Komar integrals are evaluated on the horizon and $\zeta$ is given by~\eqref{eq:mp-zeta}, this relation is an identity, and it provides a consistency check of~\eqref{eq:mp-kappa}, \eqref{eq:mp-area} and~\eqref{eq:mp-komar} rather than an independent result.
We also note that, with the normalization adopted here, these quantities do not satisfy the first law $dM_H=T_H\,dS+\Omega_H\,dJ_H$ at fixed $j$:
a consistent first law requires a careful choice of the normalization of $\partial_t$ (and possibly a thermodynamic potential conjugate to $j$), which we leave for future work.

\begin{figure}
\centering
\begin{subfigure}{0.45\textwidth}
    \centering
    \includegraphics[width=\textwidth]{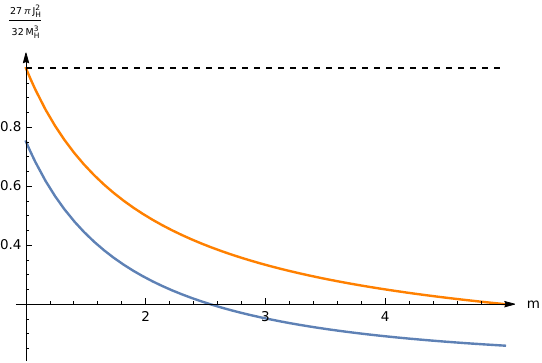}
    \caption{$a=1$, $j=0.1$.}
    \label{fig:mp-ratio1}
\end{subfigure}
\hfill
\medskip
\begin{subfigure}{0.45\textwidth}
    \centering
    \includegraphics[width=\textwidth]{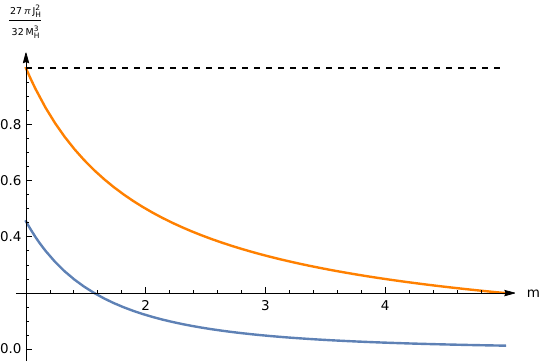}
    \caption{$a=1$, $j=0.3$.}
    \label{fig:mp-ratio2}
\end{subfigure}
\caption{Plots showing the ratio $\frac{27\pi}{32}\frac{J_H^2}{M_H^3}$ as a function of $m$ ($a$ and $j$ are fixed), for two values of $j$.
The orange lines represent the asymptotically flat Myers--Perry ratio~\eqref{eq:mp-ratio}, while the blue lines represent the swirling Myers--Perry ratio~\eqref{eq:mp-swirl-ratio}:
for these values ($aj>0$) both ratios keep their value below the horizontal dashed line with value 1.
Moreover, for $aj>0$ the swirling ratio~\eqref{eq:mp-swirl-ratio} is smaller than the standard one, and decreases with respect to~\eqref{eq:mp-ratio} by increasing the value of $j$, as one can appreciate by comparing the plots;
the opposite holds for $aj<0$.}
\label{fig:mp-ratio}
\end{figure}

In the asymptotically flat case, the values of mass and angular momentum are constrained by the inequality that guarantees the existence of the horizon, $a^2<m$.
This relation is translated, for the Myers--Perry case, into
\begin{equation}
\label{eq:mp-ratio}
\frac{27\pi}{32} \frac{J^2}{M^3} < 1 \,,
\end{equation}
that is equivalent to say that the angular momentum $J$ cannot acquire arbitrarily large values, but it is bounded above by the spacetime mass $M$.

In the swirling case, evaluation of the ratio gives
\begin{equation}
\label{eq:mp-swirl-ratio}
\frac{27\pi}{32} \frac{J_H^2}{M_H^3} = \frac{a^2}{m(1+ajm)^3} \,,
\end{equation}
so the behaviour depends on the sign of $aj$.
For $ajm\geq0$, the condition $a^2<m$ implies
\begin{equation}
\frac{27\pi}{32} \frac{J_H^2}{M_H^3} \leq \frac{a^2}{m} < 1 \,,
\end{equation}
i.e.~the bound is tighter than for the Myers--Perry black hole with the same parameters $(m,a)$, and it becomes more stringent by increasing $aj$, as shown in Fig.~\ref{fig:mp-ratio}.
For $-1<ajm<0$, instead, the ratio exceeds $a^2/m$ and is unbounded as $ajm\to-1$ (e.g.~$m=2$, $a=1$, $j=-0.4$ gives a ratio of $62.5$), so that no bound analogous to~\eqref{eq:mp-ratio} holds.
The presence of the swirling background thus creates a competition between $a$ and $j$ via the coupling of the parameters appearing in the Komar charges~\eqref{eq:mp-komar}.
We recall that $M_H$ depends on the choice of the stationary frame discussed below~\eqref{eq:omega-horizon}.

\subsection{Ergoregions}
\label{sec:mp-ergo}

We now study the existence of regions of spacetime where the Killing field $\partial/\partial t$ is spacelike, or equivalently $g_{tt}>0$, i.e.~the existence of ergoregions.
Physically, these regions are characterized by the fact that the frame dragging makes it impossible for an observer to remain static.
The boundary of the ergoregion, $g_{tt}=0$, is the ergosurface.

Inspection of the metric~\eqref{eq:mp-swirling} shows that the condition $g_{tt}>0$ is equivalent to
\begin{equation}
\Lambda|\omega| > \sin\theta\, \sqrt{\Delta} \,.
\end{equation}
The latter constraint cannot be solved analytically.
However, we notice that on the horizon, where $\Delta=0$, one has $g_{tt}=\Lambda\,\omega|_{r=r_H}^2\geq0$, which vanishes only on the axis $\theta=0$, where $\Lambda=0$:
the horizon touches the ergosurface on the $\psi$-axis.

\begin{figure}[htbp]
\centering
\begin{subfigure}[b]{0.45\textwidth}
    \centering
    \includegraphics[width=\textwidth]{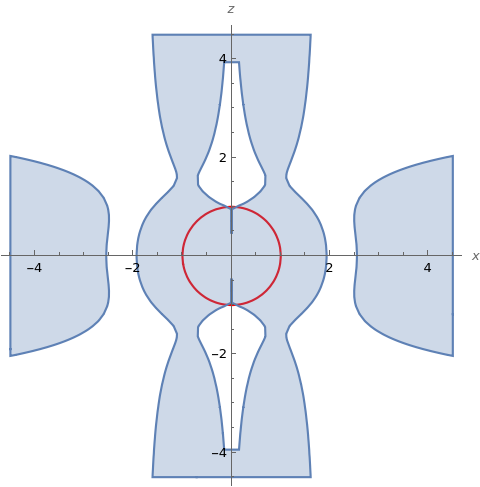}
    \caption{$m=2$, $a=1$ and $j=0.2$.}
    \label{fig:ergoj1}
\end{subfigure}
\hfill
\begin{subfigure}[b]{0.45\textwidth}
    \centering
    \includegraphics[width=\textwidth]{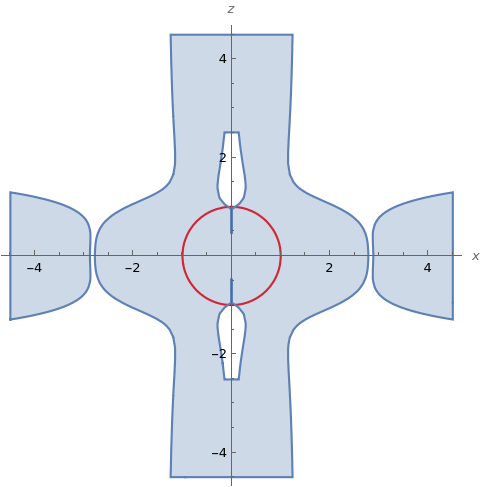}
    \caption{$m=2$, $a=1$ and $j=1$.}
    \label{fig:ergoj2}
\end{subfigure}
\medskip 
\begin{subfigure}[b]{0.45\textwidth}
    \centering
    \includegraphics[width=\textwidth]{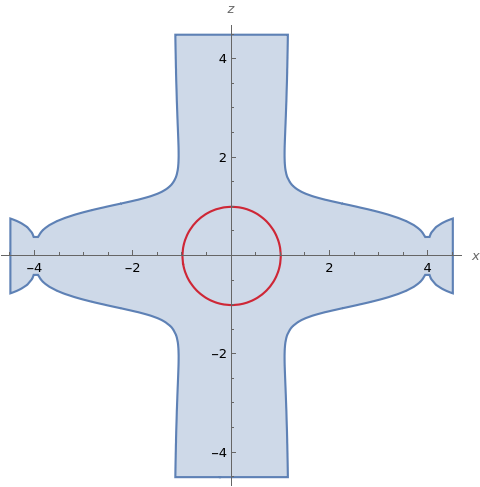}
    \caption{$m=2$, $a=1$ and $j=3$.}
    \label{fig:ergoj3}
\end{subfigure}
\hfill
\begin{subfigure}[b]{0.45\textwidth}
    \centering
    \includegraphics[width=\textwidth]{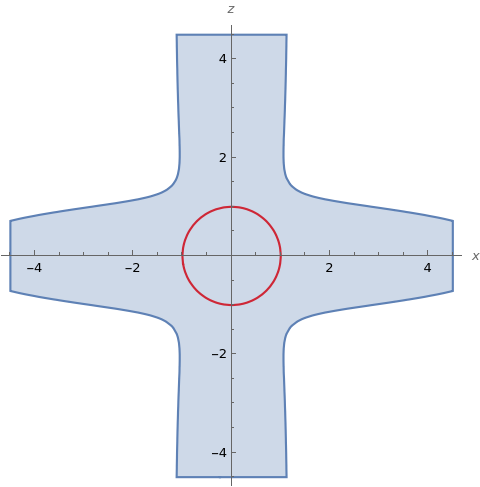}
    \caption{$m=2$, $a=1$ and $j=6$.}
    \label{fig:ergoj4}
\end{subfigure}
\caption{Ergoregion cross-sections for fixed values of $m$ and $a$, and increasing values of $j$.
The red line represents the black hole horizon.}
\label{fig:ergoOverJ}
\end{figure}

\begin{figure}[htbp]
\centering
\begin{subfigure}[b]{0.45\textwidth}
    \centering
    \includegraphics[width=\textwidth]{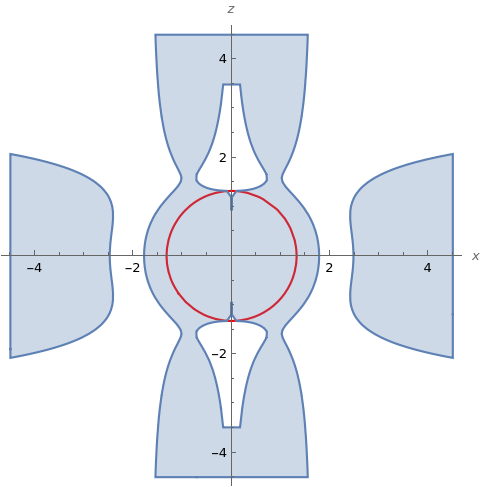}
    \caption{$m=2$, $a=0.5$ and $j=0.2$.}
    \label{fig:ergoa1}
\end{subfigure}
\hfill
\begin{subfigure}[b]{0.45\textwidth}
    \centering
    \includegraphics[width=\textwidth]{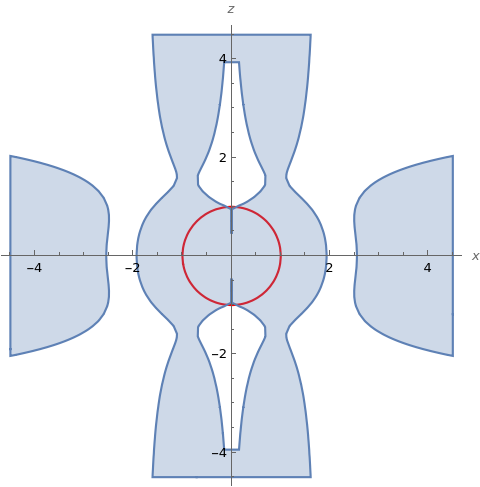}
    \caption{$m=2$, $a=1$ and $j=0.2$.}
    \label{fig:ergoa2}
\end{subfigure}
\medskip 
\begin{subfigure}[b]{0.45\textwidth}
    \centering
    \includegraphics[width=\textwidth]{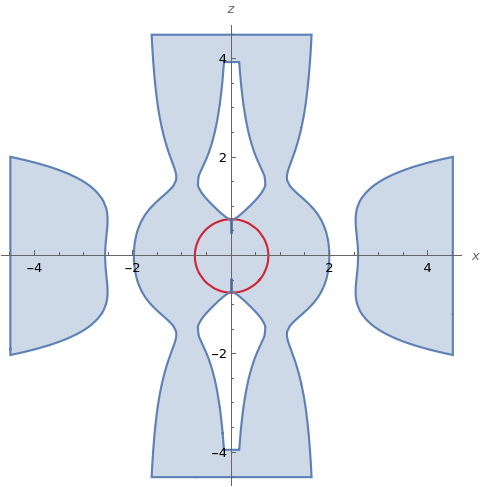}
    \caption{$m=2$, $a=1.2$ and $j=0.2$.}
    \label{fig:ergoa3}
\end{subfigure}
\hfill
\begin{subfigure}[b]{0.45\textwidth}
    \centering
    \includegraphics[width=\textwidth]{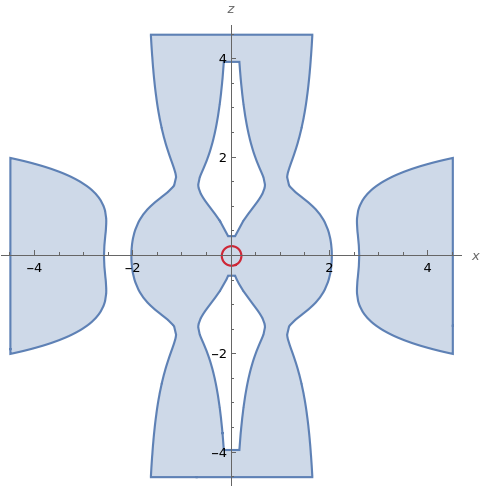}
    \caption{$m=2$, $a=1.4$ and $j=0.2$.}
    \label{fig:ergoa4}
\end{subfigure}
\caption{Ergoregion cross-sections for fixed values of $m$ and $j$, and increasing values of $a$.
The red line represents the black hole horizon.}
\label{fig:ergoOverA}
\end{figure}

We perform a numerical analysis of the inequality $g_{tt}>0$ by using rectangular coordinates~\cite{DiPinto:2025yaa}
\begin{equation}
r = \sqrt{x^2+y^2+z^2} \,, \qquad
\cos\theta = \frac{z}{\sqrt{x^2+y^2+z^2}} \,,
\end{equation}
and choosing the plane $y=0$ to represent the cross-section.
Obviously, the $\phi$-plane is suppressed in the pictorial representation.
The results of the numerical computations are shown in Fig.~\ref{fig:ergoOverJ} and~\ref{fig:ergoOverA}.

In Fig.~\ref{fig:ergoOverJ}, the four plots have fixed values of $m$ and $a$, but increasing values of $j$:
the first observation is that the horizon (depicted in red) is always surrounded by the ergoregion (depicted in blue), which extends to infinity.
This is a typical feature of swirling and Melvin spacetimes in four dimensions~\cite{Astorino:2022aam,DiPinto:2025yaa}.
For small values of $j$ the ergoregion is made of disconnected pieces, that by increasing $j$ merge to form a unique area with no ``holes'' in it, shaped as a cross.
While in the plots~\ref{fig:ergoj1} and~\ref{fig:ergoj2} the horizon intersects the ergosurface, when the swirl increases in~\ref{fig:ergoj3} and~\ref{fig:ergoj4}, the ergosurface becomes larger than the horizon.

In Fig.~\ref{fig:ergoOverA}, the plots have fixed values of $m$ and $j$, but increasing values of $a$:
the ergoregion extends to infinity, once again, while in this case the horizon evidently shrinks, because of the relation $r_H=\sqrt{m-a^2}$.
The growth of $a$ makes the ergoregion more tapered.
The disconnected pieces stay disconnected (the connectedness is related to $j$), and the ergoregion is completely symmetric (as in Fig.~\ref{fig:ergoOverJ}).
Thus, with respect to the asymptotically flat case, the ergoregion has a richer structure and evolves with $j$.
We stress that the ergoregions depend on the choice of the stationary Killing vector, i.e.~on the constant $\Omega_0$ in~\eqref{eq:mp-omega0}:
since the spacetime has no preferred asymptotic frame, a different choice $\partial_t+c\,\partial_\psi$ would change their shape.

\section{Swirling black ring}
\label{sec:ring}

We begin by reviewing the rotating black ring in the form given\footnote{Other forms of the rotating black ring can be found in~\cite{Emparan:2001wn,Emparan:2006mm,Hong:2003gx}.} in~\cite{Elvang:2003mj}
\begin{equation}
\label{eq:ring}
\begin{split}
{ds}^2 & = -\frac{F(x)}{F(y)} \bigl[ dt - R\sqrt{\lambda\nu} (1+y) d\psi \bigr]^2 \\
&\quad + \frac{R^2}{(x-y)^2} \biggl[ -F(x) \biggl( G(y) {d\psi}^2 + \frac{F(y)}{G(y)} {dy}^2 \biggr)
+ F(y)^2 \biggl( \frac{{dx}^2}{G(x)} + \frac{G(x)}{F(x)} {d\phi}^2 \biggr) \biggr] \,,
\end{split}
\end{equation}
where
\begin{equation}
F(\xi) = 1-\lambda\xi \,, \qquad
G(\xi) = (1-\xi^2) (1-\nu\xi) \,.
\end{equation}
The parameter $R$ represents the radius scale of the ring, while $\lambda$ and $\nu$ are related to the mass and angular momentum of the spacetime.
The coordinates $y$ and $x$ take values in $-1\leq x\leq1$ and $-\infty<y\leq-1$, $\lambda^{-1}<y<\infty$.
Physically, one expects that two out of three parameters are the physical ones, since in order to achieve a regular configuration the radius should be dynamically fixed by the balance between the tension and the centrifugal forces:
this is indeed the case, since~\eqref{eq:ring} is regular provided that
\begin{equation}
\label{eq:ring-conic}
\Delta\psi = \Delta\phi = 2\pi \frac{\sqrt{1+\lambda}}{1+\nu} \,, \qquad
\lambda = \frac{2\nu}{1+\nu^2} \,,
\end{equation}
where $\Delta\psi$ and $\Delta\phi$ are the periodicities of the angular coordinates.
As we notice from the latter expression, the value of one of the parameters is fixed, thus leaving us with a two-parameter solution.
We will see how this condition is modified by the presence of the swirling background.
The parameter $\nu$ is allowed to vary as $0\leq\nu<1$, while if we do not fix the parameter $\lambda$ by means of the regularity condition~\eqref{eq:ring-conic}, we take it to be $\nu<\lambda<1$ (if $\lambda\leq\nu$, a naked singularity appears);
in the limit $\nu\to0$ one recovers a non-rotating black ring, while in the limit $\nu\to1$ the ring is flattened along the rotation plane and results in a naked singularity.
Finally, $y_H=1/\nu$ is the event horizon and $|y|=\infty$ is an ergosurface.
The curvature singularity is reached as $y\to1/\lambda$ from above.

We apply the Ehlers transformation to the seed metric~\eqref{eq:ring}, to embed the rotating black ring in the swirling universe~\eqref{eq:swirl-axial}:
we define
\begin{equation}
\rho = \frac{R^2}{(x-y)^2} \sqrt{-F(x)F(y)G(x)G(y)} \,, \qquad
z = \frac{R^2}{2(x-y)^2} (1-xy) [\nu x + (\lambda+\nu)y + \lambda x(1-2\nu y) - 2] \,,
\end{equation}
and then construct the Ernst potential~\eqref{eq:ernst-potential} and map the seed metric~\eqref{eq:ring} via the Ehlers transformation~\eqref{eq:ehlers}.
The result is
\begin{equation}
\label{eq:ring-swirling}
{ds}^2 = \Lambda \bigl(d\psi-\omega dt\bigr)^2 + \Lambda^{-1} \biggl[ -\frac{\rho^2}{h^2} {dt}^2 + \frac{R^4}{(x-y)^4}\, \Xi\, (1+y) F(x) \biggl( \frac{F(x)}{G(y)} {dy}^2 - \frac{F(y)}{G(x)} {dx}^2 \biggr) \biggr] + h^2 {d\phi}^2 \,,
\end{equation}
where we defined
\begin{equation}
\Lambda^{-1} = \chi_{(0)} + j \chi_{(1)} + j^2 \chi_{(2)} \,, \qquad
\omega = \omega_{(0)} + j \omega_{(1)} + j^2 \omega_{(2)} \,,
\end{equation}
and the functions
\begin{subequations}
\begin{align}
h & = \frac{R}{x-y} F(y) \sqrt{\frac{G(x)}{F(x)}} \,, \\
\rho & = \frac{R^2}{(x-y)^2} \sqrt{-F(x)F(y)G(x)G(y)} \,, \\
\Xi & = (1-y)F(y) + \lambda\nu x^2 - \nu y \bigl[ 1 - y + \lambda(2-x)x - 2\lambda(1-x)y \bigr] \nonumber \\
& = (1-y)(1-\nu y)F(y) + \lambda\nu(1+y)(x-y)^2 \,,
\end{align}
\end{subequations}
and
\begin{subequations}
\begin{align}
\chi_{(0)} & = -\frac{(x-y)^2}{R^2} \frac{F(y)}{(1+y)F(x)\Xi} \,, \\
\chi_{(1)} & = -\frac{2R\sqrt{\lambda\nu}}{1+y} \frac{F(y)}{F(x)\Xi} \biggl[ (x-y)^2-\frac{(1+y)^2}{1+x} F(x) G(x) \biggr] \,, \\
\begin{split}
\chi_{(2)} & = -\frac{R^4(x-y)^2}{1+y} \frac{F(y)}{F(x)\Xi} \biggl[
\lambda\nu \biggl( \frac{F(x)G(x)(1-x+2y)}{(x-y)^2} - A(x) \biggr)^2 \\
&\quad + \frac{F(x)G(x)(1+y)^2 B(y,x)^2}{(x-y)^6} \biggr] \,,
\end{split}
\\
\omega_{(0)} & = \frac{(x-y)^2\sqrt{\lambda\nu}}{R\Xi} \,, \\
\begin{split}
\omega_{(1)} & = \frac{R^2}{1+y} \biggl[ \frac{F(y)G(y)}{(x-y)^2(1+y)\Xi}\Bigl( C(y) x^2 + 2 D(y) x + E(y) \Bigr) \\
&\quad + 2 + y(1+y) \bigl( \lambda + \nu + \lambda\nu(2-y) \bigr) \biggr] \,,
\end{split}
\\
\begin{split}
\omega_{(2)} & = \frac{R^5\sqrt{\lambda\nu}}{1+y} \biggl[
-\frac{F(y)G(y)}{(x-y)^4\Xi(1+y)} \Bigl( H(y) x^5 + J(y) x^4 + 2K(y) x^3 + L(y) x^2 + 2M(y) x + N(y) \Bigr) \\
&\quad + 1 + y(1+y)(\lambda +\nu) + \lambda\nu y(1+y)(1+y^2) + \lambda\nu y(1-y)(1+y)^3 (\lambda + \nu - \lambda\nu y)
\biggr] \,.
\end{split}
\end{align}
\end{subequations}
We also introduced, to shorten the expressions, the auxiliary functions
\begin{subequations}
\begin{align}
A(x) & = x \bigl[ 1 + (1-x)(\lambda + \nu - \lambda\nu x) \bigr] \,, \\
B(y,x) & = \frac{F(y)G(y) + \lambda\nu(1+y)^2(x-y)^2}{1+y} \,, \\
C(y) & = -2 + (1+y)^2 \left[ 2(\lambda+\nu) + \lambda\nu(3-5y) \right] \,, \\
D(y) & = -(\lambda+\nu)(1+3y) - \lambda\nu y - 2y^2 - \lambda y^2(1-y) + \nu y^2 \left( y - 1 + \lambda(1+3y+y^2) \right) \,, \\
E(y) & = 3 + 3y - y^2 + y^3 - y(1-y^2) \left[ \lambda + \nu - (\lambda + \nu + 2\lambda\nu)y \right] \,,
\end{align}
\end{subequations}
and
\begin{subequations}
\begin{align}
H(y) & = 2\lambda\nu(1+y)^4 \Bigl[ \lambda\big((3y-1)\nu-1\big) - \nu \Bigr] \,, \\
\begin{split}
J(y) & = \bigl((1+y)^2 \nu - 1\bigr)^2 \\
&\quad - \lambda(1+y)^2 \bigl[ \nu^2 y^3 - \nu(4\nu+5)y^2 - \nu(9\nu+13)y - 4\nu^2 + 2 \bigr] \\
&\quad - \lambda^2(1+y)^3 \bigl[ 14\nu^2 y^3 + \nu(16\nu+1)y^2 - (\nu^2+5\nu+1)y - \nu^2 - 4\nu - 1 \bigr] \,,
\end{split}
\\
\begin{split}
K(y) & = 8\lambda^2\nu^2 y^7 + 2\lambda\nu(18\lambda\nu+\lambda+\nu)y^6 + 2\lambda\nu\big(2\nu+\lambda(30\nu+2)-1\big)y^5 \\
&\quad + \big[ (44\nu^2-\nu-1)\lambda^2 - (\nu^2+14\nu+1)\lambda - \nu(\nu+1) \big] y^4 \\
&\quad + \big[ 4(3\nu^2-2\nu-1)\lambda^2 - (8\nu^2+26\nu+3)\lambda - \nu(4\nu+3) \big] y^3 \\
&\quad - \big[ (8\nu+6)\lambda^2 + (8\nu^2+20\nu+1)\lambda + 6\nu^2+\nu-2 \big] y^2 \\
&\quad + \big[ -4(\nu+1)\lambda^2 - (4\nu^2+10\nu-1)\lambda - 4\nu^2+\nu \big] y \\
&\quad - \lambda^2(\nu+1) - \lambda\nu(\nu+4) - \nu^2 \,,
\end{split}
\\
\begin{split}
L(y) & = -9\lambda^2\nu^2 y^8 - 6\lambda\nu(8\lambda\nu+\lambda+\nu)y^7 - 6\lambda\nu\big(2\nu+\lambda(15\nu+2)-1\big)y^6 \\
&\quad + 3\lambda\nu(7-24\lambda\nu)y^5 \\
&\quad + \big[ (-21\nu^2+12\nu+1)\lambda^2 + (12\nu^2+25\nu+7)\lambda + \nu^2+7\nu+1 \big] y^4 \\
&\quad + \big[ (6\nu+4)\lambda^2 + (6\nu^2+7\nu+7)\lambda + 4\nu^2+7\nu-5 \big] y^3 \\
&\quad + \big[ 6\lambda^2 + 9(\nu+1)\lambda + 6\nu^2+9\nu+3 \big] y^2 \\
&\quad + \big[ 4\lambda^2 + (16\nu+13)\lambda + 4\nu^2+13\nu+1 \big] y \\
&\quad + \lambda^2 + \nu^2 + 4\nu + 4\lambda(\nu+1) - 2 \,,
\end{split}
\\
\begin{split}
M(y) & = \lambda^2\nu^2 y^9 + 2\lambda\nu(4\lambda\nu+\lambda+\nu)y^8 + 2\lambda\nu\big(2\nu+\lambda(9\nu+2)-1\big)y^7 \\
&\quad + \lambda\nu(16\lambda\nu-5)y^6 + \big[ \nu(5\nu-4)\lambda^2 - (4\nu^2+5\nu+2)\lambda - 2\nu \big] y^5 \\
&\quad + \big[ -2\nu\lambda^2 - 2\nu^2\lambda + 3\lambda\nu + \lambda + \nu + 2 \big] y^4 + (5\lambda\nu-3)y^3 \\
&\quad - 3(2\lambda+2\nu+1)y^2 - (4\lambda+4\nu+1)y - (\lambda+\nu+1) \,,
\end{split}
\\
\begin{split}
N(y) & = 1 + 4y + 3y^2 + (1+\lambda+\nu)y^3 + (1-\lambda-\nu-3\lambda\nu)y^4 \\
&\quad - (1+\lambda+\nu+2\lambda\nu-\lambda^2\nu-\lambda\nu^2)y^5 \\
&\quad + (\lambda+\nu+2\lambda\nu+2\lambda^2\nu+2\lambda\nu^2-2\lambda^2\nu^2)y^6 \\
&\quad + 2\lambda\nu(1-3\lambda\nu)y^7 + \lambda\nu(1-2\lambda-2\nu-6\lambda\nu)y^8 \\
&\quad - \lambda\nu(\lambda+\nu+2\lambda\nu)y^9 \,.
\end{split}
\end{align}
\end{subequations}
Contrary to the Myers--Perry case presented in Sec.~\ref{sec:myers-perry}, the swirling black ring has quite involved metric components.
We will see that the intricacy of these polynomials will force us to perform some numerical computations.

The solution~\eqref{eq:ring-swirling} is characterized by the three Killing vectors $\partial/\partial t$, $\partial/\partial\psi$ and $\partial/\partial\phi$.
The position of the event horizon in the swirling black ring~\eqref{eq:ring-swirling} is left unchanged, $y_H=1/\nu$, as well as the curvature singularity in $y=1/\lambda$.
The topology of the event horizon is a deformed $S^1\times S^2$ and it rotates with respect to the axis of the Killing field $\partial/\partial\psi$.
The non-rotating black ring embedded in a swirling universe is obtained by setting $\nu=0$:
as we will see, in this case the metric is not regular.


\subsection{Regularity}
\label{sec:ring-regularity}

We are interested in establishing the regularity of the solution and check if the conical singularities are removable.
The Killing vector $\partial_\phi$ has possible conical singularities at the points $x=\pm1$, while $\partial_\psi$ at $y=-1$.

The $x\phi$ part of the metric is conformal, in the limit $x\to-1$, to
\begin{equation}
{ds}_{x\phi}^2 \propto \frac{[1-jR^3\sqrt{\lambda\nu}(1+2\nu+2\lambda(1+\nu))]^2}{2(1+\nu)s} {ds}^2 + 2\frac{1+\nu}{1+\lambda} s {d\phi}^2 \,,
\end{equation}
where $s=x+1$.
Now, define $u^2=\frac{2(1+\nu)s}{1+\lambda}$ to find
\begin{equation}
{ds}_{x\phi}^2 \propto \frac{(1+\lambda)[1-jR^3\sqrt{\lambda\nu}(1+2\nu+2\lambda(1+\nu))]^2}{(1+\nu)^2} {du}^2 + u^2 {d\phi}^2 \,.
\end{equation}
Thus, we choose the periodicity of $\phi$ as
\begin{equation}
\label{eq:ring-dphi}
\Delta\phi = 2\pi \frac{\sqrt{1+\lambda}}{1+\nu} \bigl|1-jR^3\sqrt{\lambda\nu}(1+2\nu+2\lambda(1+\nu))\bigr| \,.
\end{equation}
Now we analyze the $y\psi$ part of the metric for $y\to-1$, which is conformal to
\begin{equation}
{ds}_{y\psi}^2 \propto \frac{(1+\lambda)(1+jR^3\sqrt{\lambda\nu})^2}{2(1+\nu)s} {ds}^2 + \frac{2(1+\nu)s}{(1+jR^3\sqrt{\lambda\nu})^2} {d\psi}^2 \,,
\end{equation}
where $s=1+y$.
So we define $u^2=\frac{2(1+\nu)s}{(1+jR^3\sqrt{\lambda\nu})^2}$ to find
\begin{equation}
{ds}_{y\psi}^2 \propto \frac{(1+\lambda)(1+jR^3\sqrt{\lambda\nu})^4}{(1+\nu)^2} {du}^2 + u^2 {d\psi}^2 \,,
\end{equation}
then we choose the periodicity
\begin{equation}
\Delta\psi = 2\pi \frac{\sqrt{1+\lambda}}{1+\nu} \bigl(1+jR^3\sqrt{\lambda\nu}\bigr)^2 \,.
\label{eq:ring-dpsi}
\end{equation}
The periodicities found in the absence of the swirling background are immediately recovered when $j=0$.

Finally, we have to fix the behavior at $x=1$ for the Killing vector $\partial_\phi$:
we consider once again the $x\phi$ part of the metric, but in the limit $x\to1$, and now we find (using the periodicity $\Delta\phi$ found above)
\begin{equation}
{ds}_{x\phi}^2 \propto \frac{(1-\lambda)(1+jR^3\sqrt{\lambda\nu})^2}{(1-\nu)^2} {du}^2
+ \frac{(1+\lambda)[1-jR^3\sqrt{\lambda\nu}(1+2\nu+2\lambda(1+\nu))]^2}{(1+\nu)^2} u^2 {d\phi}^2 \,.
\end{equation}
We have to choose the parameters in such a way that the quantity
\begin{equation}
\mu = \frac{(1+\lambda)(1-\nu)^2}{(1-\lambda)(1+\nu)^2}
\biggl[\frac{1-jR^3\sqrt{\lambda\nu}(1+2\nu+2\lambda(1+\nu))}{1+jR^3\sqrt{\lambda\nu}}\biggr]^2 \,,
\end{equation}
satisfies the equation
\begin{equation}
\label{eq:conic-lambda}
\mu = 1 \,,
\end{equation}
for a certain value $\lambda=\lambda_c$.
We immediately notice that, when $j=0$, we recover the regularizing value $\lambda_c=\frac{2\nu}{1+\nu^2}$, in agreement with~\cite{Elvang:2003mj}.
On the other hand, in the non-rotating ring limit ($\nu=0$), it is impossible to achieve the equilibrium since Eq.~\eqref{eq:conic-lambda} implies $\lambda=0$, and thus we would obtain the swirling background.

\begin{figure}
\centering
\begin{subfigure}{0.45\textwidth}
    \centering
    \includegraphics[width=\textwidth]{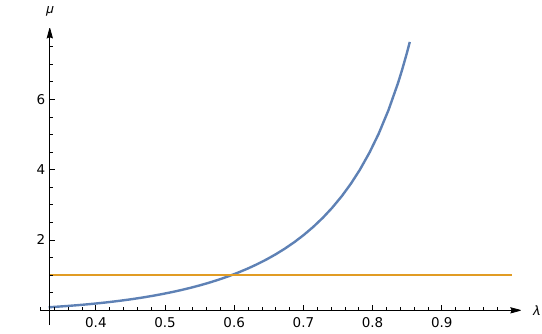}
    \caption{$\nu=1/3$, $R=1$ and $j=2$.}
    \label{fig:ring-conic1}
\end{subfigure}
\hfill
\medskip
\begin{subfigure}{0.45\textwidth}
    \centering
    \includegraphics[width=\textwidth]{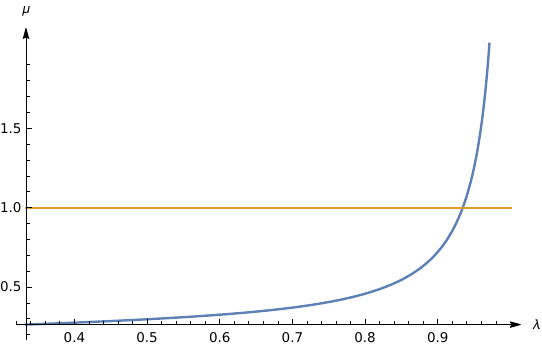}
    \caption{$\nu=1/3$, $R=1$ and $j=1/4$.}
    \label{fig:ring-conic2}
\end{subfigure}
\caption{Plots showing $\mu$ as a function of $\lambda$ (blue line), for fixed values of $\nu$, $R$ and $j$, together with the constant function 1 (orange line).
We observe that a solution $\lambda=\lambda_c$ to the Eq.~\eqref{eq:conic-lambda}, represented by the intersection between the two lines, exists.
The configuration (a) has $B<0$ and belongs to the family disconnected from the asymptotically flat black ring, while (b) has $B>0$.}
\label{fig:ring-conic}
\end{figure}

In the case of interest for us, Eq.~\eqref{eq:conic-lambda}, it is not possible to find analytically the value $\lambda_c$;
nevertheless, we can check numerically that such a value does exist.
The plots in Fig.~\ref{fig:ring-conic} show that a solution to Eq.~\eqref{eq:conic-lambda} exists, at least for some parametric ranges.

The range of existence can be characterized as follows.
Let us define
\begin{equation}
B \equiv 1-jR^3\sqrt{\lambda\nu}\bigl(1+2\nu+2\lambda(1+\nu)\bigr) \,.
\end{equation}
The solution $\lambda_c$ continuously connected to the asymptotically flat value $2\nu/(1+\nu^2)$ exists only for $\hat\jmath_-(\nu)<jR^3<\hat\jmath_+(\nu)$, with
\begin{equation}
\label{eq:ring-branch}
\hat\jmath_+ = \frac{1}{\sqrt{\nu}\,(3+4\nu)} \,, \qquad
\hat\jmath_- = -\frac{\sqrt{1+\nu}-\sqrt{1-\nu}}{\nu\bigl[\sqrt{1-\nu}\,(1+4\nu+2\nu^2)+\sqrt{1+\nu}\bigr]} \,:
\end{equation}
$\lambda_c\to1$ as $jR^3\to\hat\jmath_+$, while $\lambda_c\to\nu$ as $jR^3\to\hat\jmath_-$.
For instance, for $\nu=1/3$ one finds $-0.313<jR^3<0.400$.
On this branch $B>0$.
For $jR^3>\hat\jmath_+$ there is a second family of balanced configurations, disconnected from the asymptotically flat black ring, for which $B<0$:
in this case the absolute value in~\eqref{eq:ring-dphi} is essential.
The configuration of Fig.~\ref{fig:ring-conic1} belongs to this second family, while that of Fig.~\ref{fig:ring-conic2} lies on the branch connected to the asymptotically flat black ring.
Even if it is not possible to find an analytical expression for the regularizing $\lambda_c$, from now on we assume that such a choice has been made, and that all the quantities that we will compute are evaluated at $\lambda=\lambda_c$ to make them regular.

\subsection{Mass and angular momentum}

We now delve into the geometry of the swirling black ring~\eqref{eq:ring-swirling}.
Firstly, we compute the area of the horizon at $y=y_H$ and fixed $t$:
\begin{equation}
\begin{split}
\mathcal{A}_H & = \int_H dx\, d\psi\, d\phi \sqrt{g_{xx}g_{\psi\psi}g_{\phi\phi}} \\
& = \frac{8\pi^2R^3\sqrt{\lambda_c}(1+\lambda_c)(\lambda_c-\nu)^{3/2}}{(1-\nu)(1+\nu)^2}
\bigl(1+jR^3\sqrt{\lambda_c\nu}\bigr)^2 \bigl|1-jR^3\sqrt{\lambda_c\nu}(1+2\nu+2\lambda_c(1+\nu))\bigr| \,.
\end{split}
\end{equation}
Also in this case, the presence of the swirling background modifies the area of the horizon;
when $j=0$, the original value is recovered.

Given the definition~\eqref{eq:kappa-def}, and choosing the generator of the Killing horizon to be
\begin{equation}
\zeta = \frac{\partial}{\partial t} + \Omega_H \frac{\partial}{\partial\tilde\psi} \,,
\end{equation}
where, as in Sec.~\ref{sec:mp-smarr}, $\tilde\psi=2\pi\psi/\Delta\psi$ has period $2\pi$, and the angular velocity of the horizon is
\begin{equation}
\begin{split}
\Omega_H & = \frac{2\pi}{\Delta\psi}\, \omega \biggr|_{y=y_H} \\
& = \frac{\sqrt{\lambda_c\nu} + R^6j^2\lambda_c\sqrt{\lambda_c\nu}(1+2\nu+\lambda_c(1+\nu)(2+\nu)) + R^3 j\lambda_c(1+3\nu+2\lambda_c(1+\nu))}{R\lambda_c\sqrt{1+\lambda_c}(1+jR^3 \sqrt{\lambda_c\nu})^2} \,,
\end{split}
\end{equation}
the surface gravity~\eqref{eq:kappa-def} boils down to
\begin{equation}
\kappa_H = \frac{1-\nu}{2R\sqrt{\lambda_c(\lambda_c-\nu)}} \,,
\end{equation}
which, since it does not depend on $j$ at fixed $\lambda$, is the same result as for the asymptotically flat rotating black ring.
We have to keep in mind that all of these quantities have to be evaluated in the absence of conical singularities, i.e.~for the value of $\lambda=\lambda_c$ which satisfies Eq.~\eqref{eq:conic-lambda}.
\begin{figure}
\centering
\includegraphics[scale=1]{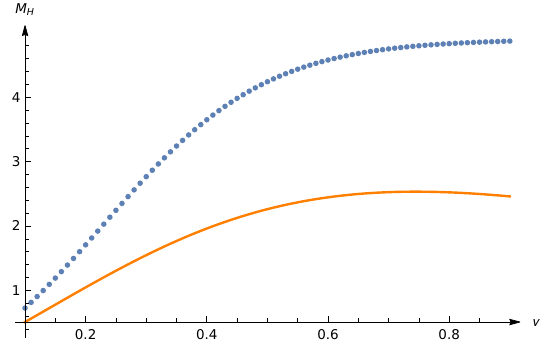}
\caption{Mass as a function of $\nu$ for the swirling black ring (blue line) and for the asymptotically flat black ring (orange line) for $R=1$ and $j=0.1$.
For this value of $j$, the mass in the swirling case is larger than the standard one.}
\label{fig:ring-mass}
\end{figure}

Mass and angular momentum are once again computed by means of Komar integrals evaluated on the horizon.
Let $\xi=\partial/\partial t$ and $\eta=\partial/\partial\tilde\psi$ be the timelike and spacelike Killing vectors, respectively, associated with time translations and rotations in the $\psi$ direction.
Then the local mass and the local angular momentum on the horizon are defined as
\begin{equation}
\label{eq:ring-komar}
M_H = -\frac{3}{32\pi} \int_H dS^{\mu\nu} \nabla_\mu\xi_\nu \,, \qquad
J_H = \frac{1}{16\pi} \int_H dS^{\mu\nu} \nabla_\mu\eta_\nu \,.
\end{equation}
In this case, however, the explicit evaluation of the Komar integrals is a formidable task, and we have not succeeded in finding an analytical form of the charges.
Again, we have to rely on a numerical evaluation of the integrals~\eqref{eq:ring-komar}.
We compute numerically the mass as a function of $\nu$ with regularity conditions, i.e.~for each value of $\nu$ we fix the corresponding value $\lambda_c$ to remove the conical singularities.
\begin{figure}
\centering
\includegraphics[scale=1]{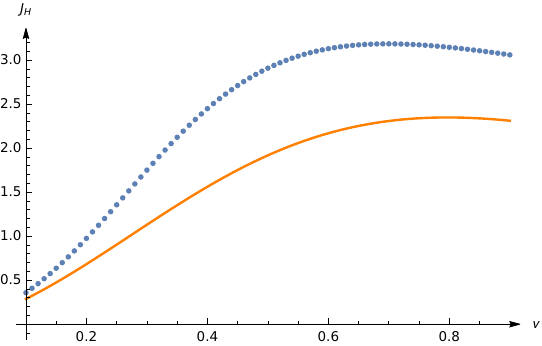}
\caption{Angular momentum as a function of $\nu$ for the swirling black ring (blue line) and for the asymptotically flat black ring (orange line) for $R=1$ and $j=0.1$.
For this value of $j$, the angular momentum in the swirling case is larger than the standard one.}
\label{fig:ring-ang}
\end{figure}

In Fig.~\ref{fig:ring-mass}, we compare the mass of the swirling black ring and the mass of the asymptotically flat black ring.
For $j>0$, we notice that the presence of the swirling parameter makes the mass larger, a behaviour that we found also in the case of the Myers--Perry black hole for $aj>0$, Eq.~\eqref{eq:mp-komar}.
A similar analysis can be performed for the angular momentum, as shown in Fig.~\ref{fig:ring-ang}:
also here, for $j>0$ the angular momentum is larger than in the asymptotically flat case.

\begin{figure}
\centering
\includegraphics[scale=1]{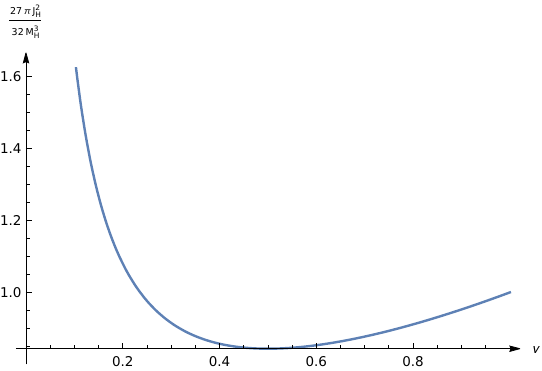}
\caption{Plot showing the ratio $\frac{27\pi}{32}\frac{J^2}{M^3}$ as a function of $\nu$ for the asymptotically flat black ring.
We see that in the range between $\frac{27}{32}$ and $1$ two asymptotically flat black rings exist with the same mass and angular momentum.}
\label{fig:ring-ratio}
\end{figure}
As in the case of the Myers--Perry black hole, we examine the value of the ratio $\frac{27\pi}{32}\frac{J_H^2}{M_H^3}$, in order to investigate the existence of constraints on mass and angular momentum.
It is known~\cite{Elvang:2003mj} that, in the asymptotically flat case, the ratio takes the form
\begin{equation}
\frac{27\pi}{32}\frac{J^2}{M^3} = \frac{(1+\nu)^3}{8\nu} \,,
\end{equation}
and it is bounded below by $27/32$, as shown in Fig.~\ref{fig:ring-ratio}.
The interesting fact is that there is a range in which the non-uniqueness of the asymptotically flat black ring is manifest, since there are two values of $\nu$ which lead to the same ratio of mass and angular momentum.
\begin{figure}[htbp]
\centering
\begin{subfigure}[b]{0.45\textwidth}
    \centering
    \includegraphics[width=\textwidth]{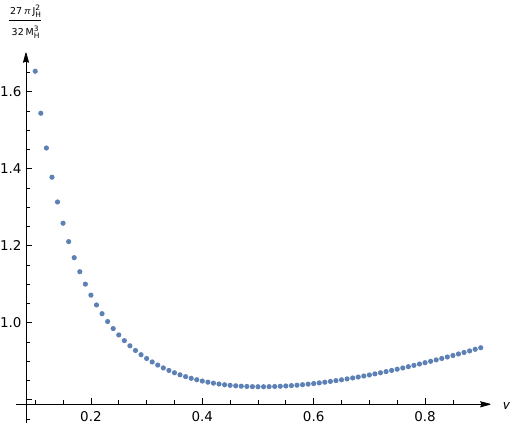}
    \caption{$R=1$ and $j=0.001$.}
    \label{fig:ring-ratio1}
\end{subfigure}
\hfill
\begin{subfigure}[b]{0.45\textwidth}
    \centering
    \includegraphics[width=\textwidth]{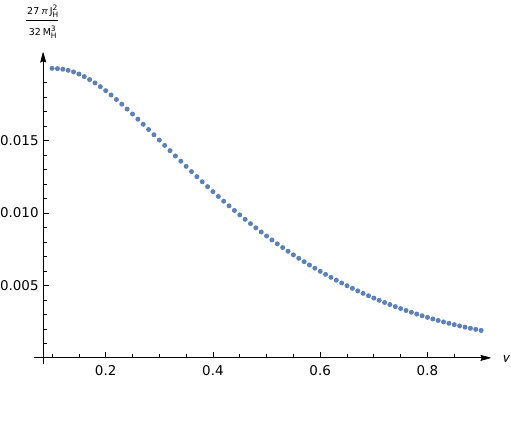}
    \caption{$R=1$ and $j=1$.}
    \label{fig:ring-ratio2}
\end{subfigure}
\medskip 
\begin{subfigure}[b]{0.45\textwidth}
    \centering
    \includegraphics[width=\textwidth]{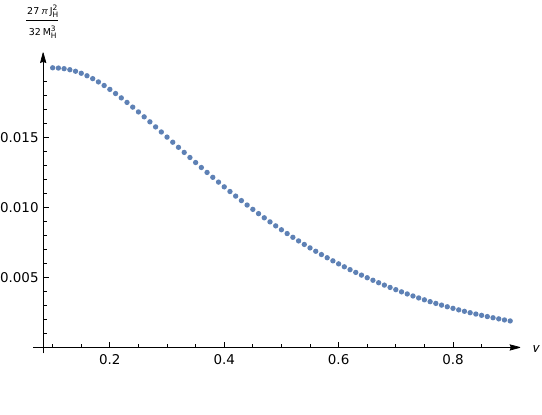}
    \caption{$R=10$ and $j=0.001$.}
    \label{fig:ring-ratio3}
\end{subfigure}
\hfill
\begin{subfigure}[b]{0.45\textwidth}
    \centering
    \includegraphics[width=\textwidth]{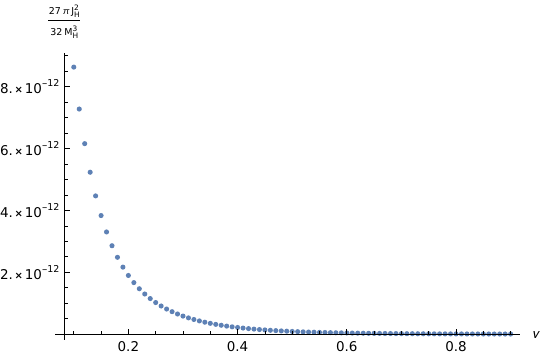}
    \caption{$R=10$ and $j=10$.}
    \label{fig:ring-ratio4}
\end{subfigure}
\caption{Plots showing the ratio $\frac{27\pi}{32}\frac{J_H^2}{M_H^3}$ as a function of $\nu$ for the swirling black ring.
The ratio depends on $R$ and $j$ only through $jR^3$, so that (b) and (c) coincide.
Configurations (b)--(d) belong to the family with $B<0$ discussed in Sec.~\ref{sec:ring-regularity}.}
\label{fig:ring-swirl-ratio}
\end{figure}

In the case of the swirling black ring, we plot in Fig.~\ref{fig:ring-swirl-ratio} the ratio versus $\nu$.
Since $j$ has dimensions of $(\text{length})^{-3}$, at fixed $\nu$ the ratio depends on $R$ and $j$ only through the dimensionless combination $jR^3$, which is why Figs.~\ref{fig:ring-ratio2} and~\ref{fig:ring-ratio3} coincide.
When $jR^3$ is small, as in Fig.~\ref{fig:ring-ratio1}, we observe a behaviour that is qualitatively similar to the asymptotically flat case (Fig.~\ref{fig:ring-ratio}).
Figs.~\ref{fig:ring-ratio2},~\ref{fig:ring-ratio3} and~\ref{fig:ring-ratio4}, with $jR^3\geq1$, correspond instead to the family of balanced rings with $B<0$ discussed in Sec.~\ref{sec:ring-regularity}, which is disconnected from the asymptotically flat black ring.
We stress that these plots do not address the question of uniqueness:
this would require comparing configurations with the same $M_H$ and $J_H$ in the same background, i.e.~at fixed $j M_H^{3/2}$, including the swirling Myers--Perry black hole of Sec.~\ref{sec:myers-perry}.
We leave this analysis for future work.

\section{Conclusions}
\label{sec:conclusions}

In this paper we used the Ehlers symmetry of the five-dimensional Ernst equation to construct swirling versions of five-dimensional Minkowski spacetime, of the Myers--Perry black hole and of the rotating black ring.
In Sec.~\ref{sec:swirl} we found that, applied to Minkowski spacetime, the map gives either the direct product~\eqref{eq:swirling} of the four-dimensional swirling universe with a line, or the genuinely five-dimensional background~\eqref{eq:swirl-axial}, which is the one relevant for the black objects;
the latter belongs to the family recently studied in~\cite{Barrientos:2026swirl}.

Then, in Sec.~\ref{sec:myers-perry} and~\ref{sec:ring} we studied the Myers--Perry black hole and the rotating black ring embedded in the swirling spacetime:
in both cases, we found that the solutions can be made free of conical singularities, and that they have no curvature singularities at finite distance outside the event horizon;
as for their background, their curvature is unbounded along the axes at large distance.
The absence of conical singularities is a remarkable feature that is not shared by the four-dimensional swirling Kerr black hole, and it is due to the fact that in five dimensions the two semi-infinite axes belong to different Killing vectors.
In the black hole case~\eqref{eq:mp-swirling}, we were able to analytically compute the horizon quantities and to check the Smarr relation;
we also found that the ratio $J_H^2/M_H^3$ is smaller than in the asymptotically flat case for $aj>0$, and larger for $aj<0$.
In the black ring case~\eqref{eq:ring-swirling}, we had to rely on numerical computations because of the intricacy of the solution:
nevertheless, we showed that the swirl modifies the balance condition of the ring, so that a branch of regular solutions connected to the asymptotically flat black ring exists only in the range~\eqref{eq:ring-branch}, while a second, disconnected family of balanced configurations appears for larger values of $jR^3$.
The question of the (non-)uniqueness of black objects in a fixed swirling background remains open.

Some directions can be taken to expand and generalize our findings:
other five-dimensional solutions can be embedded in the swirling universe~\eqref{eq:swirl-axial} by means of the Ehlers map~\eqref{eq:ehlers}.
One interesting example is given by the double-Myers--Perry solution presented in~\cite{Herdeiro:2008en}:
since in four dimensions it is known that the swirling background can help to regularize the double-Kerr spacetime~\cite{Astorino:2025zse}, a similar mechanism may work also in higher dimensions.
In our case, the hope is strengthened by the fact that even the single black hole case is free of conical singularities, as shown in Sec.~\ref{sec:myers-perry}.
Another example is provided by the lens-space black hole~\cite{Chen:2008fa}, that is affected by a conical singularity:
one can check if the swirling background removes such a defect.

One may wonder if there exists a higher-dimensional generalization of our solutions.
The generalization of the swirling background~\eqref{eq:swirling} in Weyl coordinates to an arbitrary number of dimensions is immediate:
for a $D$-dimensional spacetime
\begin{equation}
\label{eq:swirl-higher}
{ds}^2 = (1+j^2\rho^4) \bigl( -{dt}^2 + {d\rho}^2 + {dz}^2 \bigr)
+ \frac{\rho^2}{1+j^2\rho^4} (d\psi - 4jz\, dt)^2 + \sum_{i=1}^N {d\phi}_i^2 \,,
\end{equation}
where $N=D-4$.
Since the extra dimensions are inert, the qualitative behaviour of such a spacetime is the same that we analyzed in Sec.~\ref{sec:swirl}.
A more interesting question is the following:
is it possible to construct a $D$-dimensional swirling spacetime with $\lfloor\frac{D-1}{2}\rfloor$ independent swirling parameters $j_i$?
In five dimensions, a second independent swirling parameter has been obtained in~\cite{Barrientos:2026swirl} by using the full coset formulation, where it is also shown that, within the same Ehlers construction, at most two rotational swirls can be activated on a globally defined background.
Other approaches may be given by the study of a matrix Ernst equation, along the lines of~\cite{Alekseev:2004zz}, in such a way to add independent Ehlers parameters to the seed solution (in the case in which a generalized Ernst transformation is still a symmetry of the equation).
Another interesting point is the embedding of black objects in the background~\eqref{eq:swirl-higher} or in its multi-parameter generalizations.
Black strings and branes in~\eqref{eq:swirl-higher} are trivially obtained as products of the four-dimensional swirling black holes~\cite{Astorino:2022aam} with flat directions, while localized black holes would require a solution generating technique;
once again, a matrix Ernst model might be the right approach.

The extension to the five-dimensional Einstein--Maxwell system can be easily treated, at least in the setting of~\cite{Yazadjiev:2008pt}:
in that case, the Ehlers map~\eqref{eq:ehlers} is still an exact symmetry of the equations of motion and one can still take advantage of it to embed charged solutions in a swirling background (e.g.~constructing a swirling Reissner--Nordstr\"om black hole or a Melvin-swirling universe as in~\cite{DiPinto:2025yaa}).

A five-dimensional vacuum solution can be easily cast into a four-dimensional Einstein--Maxwell-dilaton solution, as it is well known~\cite{Kaluza:1921tu,Klein:1926tv}.
In this respect, it might be interesting to study the four-dimensional solutions that can be obtained from the exact spacetimes found in Secs.~\ref{sec:swirl},~\ref{sec:myers-perry} and~\ref{sec:ring}.
Reduction along $\partial/\partial\phi$, which is hypersurface-orthogonal, gives four-dimensional Einstein-dilaton solutions, while reduction along $\partial/\partial\psi$ gives Einstein--Maxwell-dilaton configurations;
in both cases, the four-dimensional fields are singular where the Killing vector used in the reduction has fixed points.

\section*{Acknowledgements}
This work was partly supported by INFN.

\end{document}